\documentclass[letter]{aa} 
\usepackage{graphicx}
\usepackage{txfonts}
\usepackage{natbib}
\begin{document}
\title{Low-mass lithium-rich AGB stars in the Galactic bulge: evidence for
  Cool Bottom Processing?\thanks{Bas\-ed on ob\-ser\-va\-tions at the Ve\-ry
  Lar\-ge Te\-le\-scope of the Eu\-ro\-pe\-an South\-ern Ob\-ser\-va\-to\-ry,
  Cer\-ro Pa\-ra\-nal/Chi\-le under Program 65.L-0317(A, B).}}

\author
    {
      S. Uttenthaler\inst{1}
      \and
      T. Lebzelter\inst{1}
      \and
      S. Palmerini\inst{2}
      \and
      M. Busso\inst{2}
      \and
      B. Aringer\inst{1}
      \and
      M. T. Lederer\inst{1}
    }

\offprints{S. Uttenthaler, uttenthaler@astro.univie.ac.at}

\institute
    {
      Department of Astronomy, University of Vienna,
      T\"ur\-ken\-schanz\-stra\ss e 17, 1180 Vienna, Austria\\
      \email{(uttenthaler;lebzelter;aringer;lederer)@astro.univie.ac.at}
      \and
      Department of Physics, University of Perugia, Via A. Pascoli 1,
      06123 Perugia, Italy\\
      \email{palmerini;busso@fisica.unipg.it}
    }

\date{Received May 14, 2007; accepted June 26, 2007}


\abstract
{The stellar production of the light element lithium is still a matter of
  debate.}
{We report the detection of low-mass, Li-rich Asymptotic Giant Branch (AGB)
  stars located in the Galactic bulge.}
{A homogeneous and well-selected sample of low mass, oxygen-rich AGB
  stars in the Galactic bulge has been searched for the absorption lines of
  Li. Using spectral synthesis techniques, we determine from high resolution
  UVES/VLT spectra the Li abundance in four out of 27 sample stars, and an
  upper limit for the remaining stars.
}
{Two stars in our sample have a solar Li abundance or above; these stars seem
  to be a novelty, since they do not show any s-element enhancement. Two more
  stars have a Li abundance slightly below solar; these stars do show
  s-element enhancement in their spectra. Different scenarios which lead to an
  increased Li surface abundance in AGB stars are discussed.}
{Of the different enrichment scenarios presented, Cool Bottom Processing (CBP)
  is the most likely one for the Li-rich objects identified
  here. Self-enrichment by Hot Bottom Burning (HBB) seems very unlikely as all
  Li-rich stars are below the HBB mass limit. Also, the ingestion of a low
  mass companion into the stars' envelope is unlikely because the associated
  additional effects are lacking. Mass transfer from a former massive binary
  companion is a possible scenario, if the companion produced little s-process
  elements. A simple theoretical estimation for the Li abundance due to CBP is
  presented and compared to the observed values.}

\keywords{Physical processes: nucleosynthesis -- Stars: AGB and post-AGB --
  Stars: evolution -- Stars: abundances}

\maketitle


\section{Introduction}

The AGB represents the final nuclear-burning phase of stellar evolution for
stars with initial masses between $\sim$~0.8 and 8.0\,M$_{\sun}$. The internal
structure consists of an electron-degenerate carbon-oxygen core which is the
remainder of He-core and -shell burning, a He-rich shell on top of which H is
burning in a shell, and a deep outer convective envelope. For a comprehensive
review of AGB evolution see e.g. \citet{HabingOlofsson}. In the most luminous
part of the AGB, the behaviour of a star is characterised by the so-called
thermal pulses (TP), thermal instabilities of the He shell accompanied by
changes in luminosity, temperature, period, and internal structure
\citep{Busso99,Herwig}. Before a star reaches the AGB its previous evolution
has significantly changed its surface abundance pattern. For instance, the
$^{12}$C/$^{13}$C ratio has been considerably lowered from its initial (solar)
value by the first dredge-up on the lower RGB. A remarkable effect is the
diminishing of Li. This element is destroyed in stars once it is mixed to
layers where the temperature exceeds $3.0 \times 10^6$\,K
\citep{Bodenheimer}. Due to its fragility, lithium is an important diagnostic
tool for stellar evolution \citep{Rebolo}, but also has a high significance in
the determination of cosmological parameters \citep{Spites,Korn}.

In intermediate mass ($\gtrsim 4 - 8$~M$_{\sun}$) thermally pulsing AGB
(TP-AGB) stars, the outer convective envelope penetrates into the H-burning
shell, thus nuclear-burning occurs partly under convective conditions. This
process is generally known as Hot Bottom Burning \citep[HBB;
][]{Iben1973}. Under these conditions, $^7$Li is produced via the so-called
Cameron-Fowler or $^7$Be-transport mechanism \citep[$^3$He($\alpha$,
  $\gamma$)$^7$Be($e^-$, $\nu$)$^7$Li; ][]{CameronFowler}. The first detection
of massive Li-rich AGB stars was reported by \citet{SL1989,SL1990}, and
confirmed later to be in agreement with HBB models \citep{Smith95}. Most
recently, \citet{GarciaHernandez} presented Li abundance measurements of
massive Galactic AGB stars.

Different types of evolved stars with Li not fulfilling the criteria for HBB
are known in the literature. Li-rich K-type stars are presented in
\citet{delaReza1997}. A Li-rich red giant (probably RGB) has been reported by
\citet{Smith1999}. Only a few low luminosity S-type \citep{Smith95, vaneck}
and C-type \citep{Abia} Li-rich AGB stars are known.

In this paper we present the discovery of Li-rich, low mass M-type AGB
stars in the Galactic bulge and discuss different mechanisms for the Li
enrichment.

\section{Observations and Analysis}

Spectra of the present sample of O-rich (M-type) Galactic bulge AGB stars have
been originally obtained for a search for the radioactive element technetium
(Tc) in these stars, which is an indicator of recent or ongoing third
dredge-up (3DUP). The observed spectra and their analysis with respect to the
occurrence of lines of Tc has been presented in \citet[][hereafter called
  Tc-paper]{Uttenthaler2007} in detail.

The homogeneous and well-selected sample of long-period variables in the
Palomar Groningen field no.~3 (PG3), located in the outer Galactic bulge
(Galactic latitude $b \sim -10^{\circ}$), has been collected from
\citet{Wess87}. Twenty-seven targets out of this sample were observed with the
UVES spectrograph at ESO's VLT in July 2000. Besides the blue arm
(377$-$490~nm) that was used to search for the occurrence of Tc, also the red
arm (667$-$847~nm and 865$-$1050~nm) was covered by the observations. The
signal-to-noise ratio (SNR) -- as given by the ESO pipeline -- achieved in the
spectral region of the Li resonance doublet varies quite strongly between 14
and 240 per pixel, although most stars (including the ones reported here to
show Li) have an SNR greater than 30.

From the occurrence of Tc in four out of the 27 sample stars, and from their
periods and luminosities, a minimum main sequence mass of
$\sim$1.5\,M$_{\sun}$ for the Tc-rich stars was estimated. This has an effect
on the inferred bulge age, since stars of that mass cannot be much older than
3$-$4 Gyrs. This is in contrast to what is found from studies of other stellar
types in the bulge, but well in line with a number of works on bulge AGB stars
(see Tc-paper).

To determine the stellar parameters of our stars for the abundance analysis,
we calculated a small grid of hydrostatic model atmospheres. The atmospheres
are based on the MARCS code (see Tc-paper for details). The grid covered the
following stellar parameter space: $T_{\mathrm{eff}}$~=~2600 $-$ 3600~K in
steps of 100~K, $\log(g) = -0.5$ and 0.0, [M/H]~=~$-0.5$, 0.0, and +0.2. Ti
was enhanced by 0.2\,dex in the models \citep{McWilliamRich}. Due to their
rather limited effect on the appearance of the spectra the mass and the C/O
ratio were fixed to the values 1.0~M$_{\sun}$ and 0.48, respectively. The
microturbulent velocity was set to 3~kms$^{-1}$.

Synthetic spectra based on these model atmospheres were calculated for the two
wavelength ranges 668 $-$ 674~nm and 700 $-$ 710~nm, respectively. The first
piece covers (besides the Li line) the TiO $\gamma$(1,0)Rc and $\gamma$(1,0)Rb
band heads, the second piece covers the TiO $\gamma$(0,0)Ra and
$\gamma$(0,0)Rb band heads. These band heads are rather sensitive to
$T_{\mathrm{eff}}$. The synthetic spectra were convolved with a Gaussian to
reduce the resolution to the value of the observed spectra (R~=~50\,000), and
an additional macro-turbulence of 4 kms$^{-1}$ was added in the convolution.

A $\chi^2$ minimisation method over the mentioned spectral ranges was applied
to find the main parameters of the sample stars. As comparison between the two
spectral regions used for the fit showed, the temperature yielding the best
fit differed substantially between the two regions in many of the observed
stars. On average, the temperature derived from the region around the Li-line
was lower by about 100~K then the temperature derived from the second
wavelength range. We suspect the reason for this is an underestimation of the
TiO $\gamma$(1,0)Rc and $\gamma$(1,0)Rb band head strength in the line list
used \citep{Schwenke}. A similar problem has been reported in \citet{Reiners}
for other band heads of TiO. Thus, it seems possible that several TiO bands
are incorrect in their strength, and a revision of the current line list is
desirable.

Spectral synthesis calculations were applied in order to identify stars which
show signs of Li line absorption. For these calculations we chose the model
yielding the lowest $\chi^2$ value in the spectral region around the Li line,
despite the obvious discrepancy regarding the derived temperature. The reason
for this is that the background TiO absorption is thus modelled as good as
possible. The four stars with a positive Li detection could all be fitted with
a single atmospheric model. This model has $T_{\mathrm{eff}}$~=~3000~K,
$\log(g) = 0.0$, and [M/H]~=~0.0.

\begin{table}
\caption{Sample stars with positive Li detection. The bolometric magnitude is
  corrected for depth effects within the bulge.}
\label{Liabund}
\begin{tabular}{lrrccr}
\hline\hline
Name & $M_{\mathrm{bol}}$ & P (days) & $\log \epsilon(\mathrm{Li})$ & $\Delta
\log \epsilon(\mathrm{Li})$ & Tc? \\
\hline
\object{M45}   & $-$4.52 & 271.02 & 2.0 & 0.5 & no  \\
\object{M794}  & $-$4.76 & 303.54 & 1.1 & 0.4 & no  \\
\object{M1147} & $-$5.28 & 395.63 & 0.8 & 0.4 & yes \\
\object{M1347} & $-$5.43 & 426.60 & 0.8 & 0.4 & yes \\
\hline
\end{tabular}
\end{table}

We assessed the content of s-process elements in our sample stars by
investigating the strength of the ZrO band head at 462\,nm with respect to the
TiO $\gamma$(0,0)Ra band head at 705.6~nm. Also, atomic line strengths of the
s-elements Sr, Y, Zr, Ba, La, Ce, Nd, and Sm in the range 400 -- 460~nm were
inspected. Only the previously identified Tc-rich stars showed an increased
strength of these features.

\section{Discussion}

\begin{figure}
  \centering
  \includegraphics[width=8.5cm]{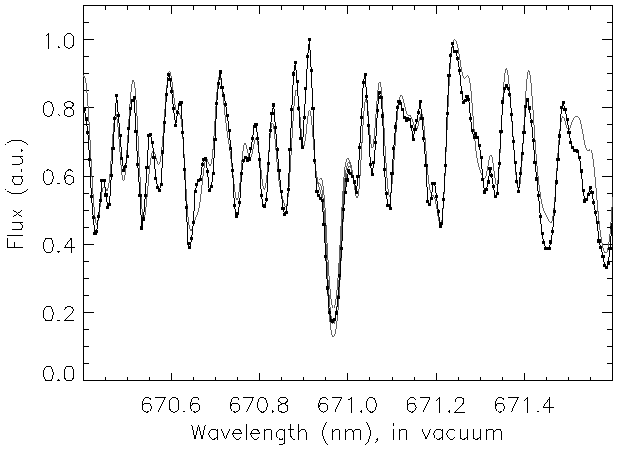}  
  \caption{
    Observed spectrum of the most Li-rich sample star M45 around the 671~nm Li
    line together with the synthetic spectrum (grey line) used for the Li
    abundance determination, assuming Li abundances of $\log
    \epsilon(\mathrm{Li}) = 1.8$, $\log \epsilon(\mathrm{Li}) = 2.0$, and
    $\log \epsilon(\mathrm{Li}) = 2.2$.}
  \label{M45Li}
\end{figure}

For four stars a non-negligible Li abundance has to be assumed in order to
give satisfying fits to the observed spectrum. Their determined Li abundances
and bolometric magnitudes (corrected for the bulge depth scatter, see the
Tc-paper) are summarised in Table~\ref{Liabund}. The error on the Li
abundance given in the fifth column was estimated by varying the temperature of
the atmospheric model by $\pm 100$~K. For the other stars, an upper limit to
the Li abundance was estimated. For the hotter stars ($\sim$3400\,K,
Semi-regular variables), this upper limit is around $\log
\epsilon(\mathrm{Li}) = 0.6$, for the cooler stars ($\sim$3000\,K, Mira
variables), it is around $\log \epsilon(\mathrm{Li}) =
0.1$\footnote{Throughout the paper, the Li abundance is given on the scale
  $\log \epsilon(\mathrm{Li}) = \log N(\mathrm{Li})/N(\mathrm{H}) + 12$. The
  solar photospheric Li abundance on this scale is $\log \epsilon(\mathrm{Li})
  = 1.1$.}. Actually, excellent fits for the spectra of the Li-poor stars can
be achieved by fully neglecting Li in the spectral synthesis. Fig.~\ref{M45Li}
shows the observed spectrum of M45, the most Li-rich star in our sample, along
with synthetic spectra with differing Li abundance.

The first thing we note is that the bolometric magnitudes of the stars listed
in Table~\ref{Liabund} are considerably below the luminosity expected for
Li-enrichment due to HBB. [In models with rather extreme convective overshoot,
  HBB may start already at luminosities as low as $M_{\mathrm{bol}} \simeq
  -5\fm0$ \citep{thesiskarakas}, but Li-enrichment becomes observable only at
  $M_{\mathrm{bol}} \simeq -6\fm0$ \citep{Smith95,Vanture}]. Regarding the
bolometric magnitudes of these stars, and also age considerations of the
Galactic bulge, the HBB scenario is very unlikely to account for the observed
Li abundances.

A few stars with Li enrichment but $M_{\mathrm{bol}}$ below $-$6\fm0 have been
reported in the literature. \citet{Smith95} identify a low luminosity, S-type
star with very high Li abundance in the SMC (\object{HV 1645}). They state a
bolometric magnitude of $-$4\fm68 for that star. \citet{Abia} derives
$M_{\mathrm{bol}}=-5\fm0$ for the C-rich star WX Cyg. Three S-stars below the
HBB limit, namely \object{NO Aur}, \object{$\pi^{1}$ Gru} and \object{HR Peg},
are listed by \citet{vaneck}.

Another remarkable thing about M45 and M794 in our sample is the absence of
Tc-lines from their spectra. Both stars do not show any other signs of
s-process enhancement either, while M1147 and M1347 do show Tc and s-process
enhancement. Among a sample of Galactic MS- and S-type stars \citet{Vanture}
found three groups of stars with respect to the Li and Tc abundance: those
without Tc and Li, interpreted as the result of mass transfer from a more
massive companion; those with Tc and Li, which are intermediate-mass stars
producing Li through HBB; and those with Tc but no Li, which are thought to be
AGB stars below the HBB mass limit. One star of their sample, \object{V441
  Cyg}, shows no Tc but is enriched in Li. This star as well as the low
luminosity S- and C-type stars with Li found before were interpreted as due to
Cool Bottom Processing (see Sect.~\ref{SectCBP}).

The stars M45 and M794 in our sample seem to be indeed a novelty, since they
are the first AGB stars detected with a considerable Li overabundance, but no
indications for a third dredge-up. From their variability and their location
in the bulge CMD both stars are certainly on the AGB. In the following we will
discuss these two stars in the light of the various Li-enrichment scenarios.

\subsection{Enrichment by massive binary companion}

The Li-rich stars might have received Li-rich matter by wind accretion from a
close massive ($M \gtrsim 4$~M$_{\sun}$) binary companion which experienced
HBB and might be a white dwarf now. One might think that a possible mass
transfer would have left its signature not only in the presence of Li, but
also in the abundance of s-process elements, since such a massive star can be
expected to have undergone s-processing and 3DUP during its TP-AGB phase. This
mass transfer would not have shown up in our search for Tc due to the short
life time of this element. As mentioned above, our Li-only stars show also no
enhancement in the stable s-process elements. However, Li enrichment is
expected to occur in early AGB phases \citep{thesiskarakas}, when s-elements
might still be absent. Moreover, intermediate mass stars might never become
very rich in s-elements, as their prevailing $^{22}$Ne source never builds up
large neutron exposures, and the massive envelope dilutes extensively any new
material dredged-up. Indeed, \citet{GarciaHernandez} did not find enhanced ZrO
band and atomic Zr~I, Nd~II, and Ba~II line strengths in their sample of
massive Galactic Li-rich AGB stars. This implies only a low s-process level in
these stars. In contrast, \citet{Smith95} find a considerable level of
s-process enrichment for their Li-rich AGB stars in the Magellanic Clouds. The
difference in the behaviour of massive AGB stars located in the Magellanic
Clouds and in the Milky Way galaxy might be a metallicity effect, but
conclusive interpretation is not yet possible. It is not known how the bulge
fits into this picture.

A possible WD companion may be detectable via an excess flux in the U or B
band. Mean B and R band values for the two most Li-rich stars, M45 and M794,
were found in the literature, but they did not turn out to be conspicuously
blue in $(B - R)$.

In conclusion, the massive binary companion hypothesis is not fully
convincing, but cannot be ruled out on the basis of current observations.

\subsection{Accretion of a (sub-)stellar companion}

\citet{SiessLivio} investigated the response of the structure and abundances of
an AGB star to the accretion (``swallowing'') of a (sub-)stellar companion.
One result of their considerations is that the effect on the Li abundance
might only be detectable if a considerable mass ($\gtrsim 0.1$~M$_{\sun}$) is
accreted to the envelope of the AGB star. Other expected effects of a brown
dwarf accretion include the increase of the mass loss rate and a spin-up of
the envelope. These would be detectable as an IR excess and the rotational
broadening of spectral lines, respectively. M45 and M794, the two most Li-rich
stars in our sample were not detected with IRAS, while M1147 and M1347 were.
For the latter two mass loss rates of the order of $10^{-6.6}$~M$_{\sun}$/yr
have been derived in our Tc-paper. Since these two stars are also the
brightest and longest period stars in our sample, this mass loss rate can be
expected due to normal evolution. Also, \citet{SiessLivio} estimate the
fraction of stars with a low-mass companion which may be accreted during the
evolution of the primary to 4$-$8 percent. However, 15\% of our sample stars
show Li. We conclude that the sub-stellar accretion scenario is unlikely to
explain the Li abundances observed in our sample stars.

\subsection{Cool Bottom Processing}\label{SectCBP}

It is now known \citep[see e.g.][]{kra94,Char04} that the radiative layers
below the convective envelope in evolved, low-mass red giants are the site of
slow mixing phenomena, in addition to the convective dredge-up episodes. Such
phenomena have been variously called ``deep mixing'', or ``extended mixing'',
or {\it Cool Bottom Processing} (CBP), although this last name is relatively
recent \citep{Wasserburg1995}. The chemical stratification of a radiative
layer hampers mass motions; hence, mixing can take place only after the
H-burning shell has erased any chemical discontinuity left by core H burning
and by the first dredge-up. When this occurs, the luminosity shows a bump,
after which abundance changes related to extra-mixing begin to appear.

As to the physical origin of the extended mixing phenomena, very recently a
number of hypotheses have been presented, at least for RGB stars.
\citet{Eggleton}, on the basis of a 3D simulation, found that Rayleigh-Taylor
instabilities below the convective envelope can develop due to the inversion
of the mean molecular weight gradient induced by $^{3}$He
burning. Alternatively, \citet{ChaZah} suggested that the double-diffusive
mechanism called ``thermohaline instability'' should be at play. In principle,
both these phenomena might occur also on the AGB, though detailed models have
not been presented. Finally, \citet{Busso07} explore the possibility that
circulation of partially processed matter can be accounted for by magnetic
buoyancy induced by a stellar dynamo operating on the RGB and on the AGB.

Independently of the still uncertain physical cause, we know that the
best established effect is a shift in the carbon isotopic mix \citep{gil91},
decreasing from the typical value of 25$-$30 left by the first dredge-up to
10$-$15 for Population I stars and down to 4 for Population II objects. Other
consequences include N enrichment and $^{18}$O depletion
\citep{chana}. Similar CBP episodes can occur later, during the thermally
pulsing AGB stages, where again the sub-adiabatic zone below the envelope has
a homogeneous molecular weight. Several authors \citep{bs99, charbal} studied
the possibility that $^7$Li is affected by circulation phenomena. In
particular it was noted that, on the RGB, Li enrichment is limited to phases
close to the bump itself, so that it was suggested that Li production
accompanies the early onset of extra-mixing \citep{charbal}.

Let us illustrate what is found in stellar codes, using a model star of
initially 1.5~M$_{\sun}$, with a Population I composition ([Fe/H]~=~$-0.3$;
Busso et al., 2003; Straniero et al., 2003).
Here production of $^7$Li derives entirely from (bound or free) electron
captures on $^7$Be. The synthesis of $^7$Be can be followed through the
competition of production and destruction:

$$
{dN(^7Be) \over dt} = N(3) N(4) \lambda_{3,4} - N(^7Be) N(p)
\lambda_{7,p} - N(^7Be) \lambda_{7,e^-} \eqno (1)
$$

\noindent where N(3), N(4), N(p) are the number abundances of $^3$He, $^4$He
and protons, and $\lambda$ indicates the reaction rate. In high temperature
regions ($T > 2 \times 10^7$~K) $^7$Be is completely ionised so that the
contribution to e$^-$ captures coming from bound electrons is suppressed. Here
the synthesis of $^8$B through p captures is efficient and prevails.

\begin{figure}
 \centering
 \includegraphics[width=8.5cm]{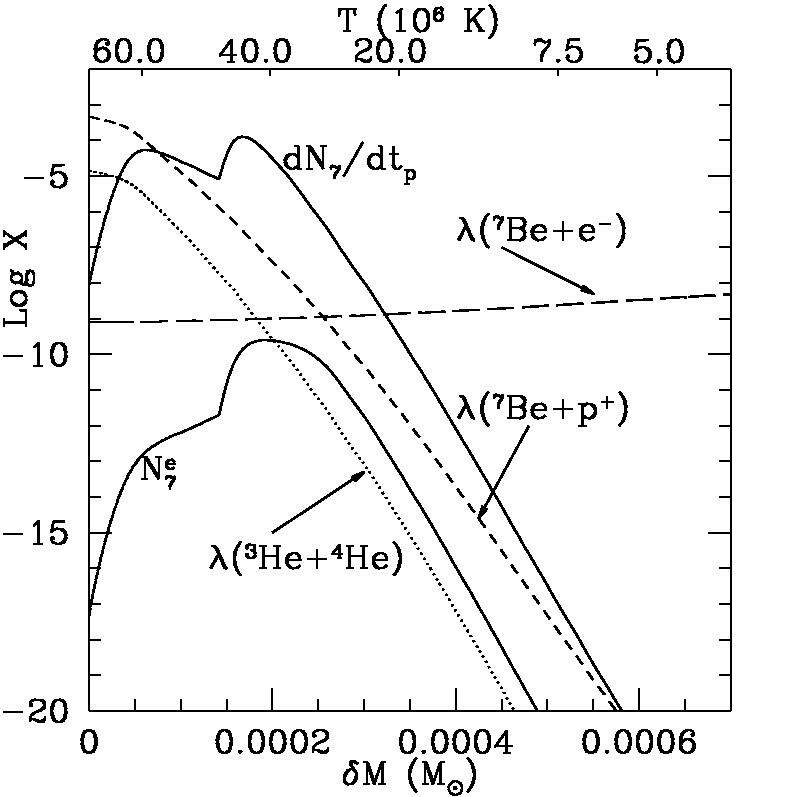}  
 \caption{Rates for $^7$Be production and destruction in the radiative layers
 below the convective envelope (dotted and dashed lines). The abscissa shows
 the distance in mass from the deepest layers affected by CBP, so that the
 convective envelope is at the right, outside the plot area. Also plotted is
 the $^7$Be equilibrium abundance ($N_7^e$) and its production rate (without
 destruction) if an infinite $^3$He reservoir is available $(dN_7/dt)_p$. This
 mimics the supply from the $^3$He present in the envelope in case of extended
 mixing.
 }
 \label{cbp}
\end{figure}

In the mentioned stellar model the mass dependence of the rates for $^7$Be
production [$\lambda(^{3}He + ^{4}He)$] and for its destruction through proton
[$\lambda(^{7}Be + p^{+})$] or electron [$\lambda(^{7}Be + e^{-})$] captures
is shown in Fig.~\ref{cbp}, for the case where p-captures occur without mass
circulation. In this case $^7$Be reaches an equilibrium concentration $N_7^e$
that depends on the $^3$He abundance maintained by H burning, and that remains
pretty low. In the same hot zones, and down to $T \ge 3-4 \times 10^6$~K, any
$^7$Li remaining captures protons and is very efficiently destroyed.

The same qualitative behaviour is maintained in presence of circulation
phenomena too slow to save most of the produced $^7$Be into regions of low T
(where it decays into $^7$Li). Actually, any mixing phenomenon occurring at
sufficiently low rates (a few 10$^{-8}$ $M_{\sun}$/yr characterises Red
Giants) would destroy Li in the envelope, carrying it to hot regions where it
is burnt, while its replenishment through $^7$Be saved at low T would remain
too small to compensate.

In order to achieve a net and durable Li enhancement in the envelope it is
required that a mixing mechanism of the type inferred by \citet{CameronFowler}
carries $^7$Be rather rapidly out of the high-temperature zone, typically to
$< 3 \times 10^6$~K. On the one hand in such low-T regions the ionisation
equilibrium of $^7$Be favours the presence of bound electrons, from which
$e^-$ captures are increased. These dominate over p-captures (that are
essentially shut off). On the other hand, in these layers any $^7$Li produced
would survive and the total inventory of $^7$Li to be carried to the envelope
can be largely increased. We can have an estimate for the maximum $^7$Li
production if we consider the pure production of $^7$Be in equation (1),
without destruction by p-captures, as if it were quickly carried to cool zones.
Its derivative in time is shown in Fig.~\ref{cbp} as $(dN_7/dt)_p$. The
integral of that curve over the production region and over a period of one
year provides the {\it maximum} mass of Li that can be produced per year,
amounting to $5.65 \times 10^{-13} M_{\sun}/yr$.
In order to have the envelope (of about 1~M$_{\sun}$) enriched up to a mass
fraction X(Li)~=~$2 \times 10^{-8}$ ($\simeq 10$ times solar) we need to mix
the produced Li at a rate $\dot M > 3 \times 10^{-5}$~M$_{\sun}$/yr, which is
a rather fast circulation rate. Such high mixing rates have been so far
suggested to occur only in rather extreme AGB cases, in order to explain some
isotopic and elemental shifts there inferred from measurements on pre-solar
grains of AGB origin. They would produce isotopic and elemental abundance
changes (including Li) hardly distinguishable from hot bottom burning effects
\citep{Nollett}, but for much lower stellar masses.
In previous calculations with complete models \citep[e.g. ][]{bs99} Li
production was obtained only adopting very fast circulation rates ($\dot M
\simeq 10^{-4} \mathrm{M}_{\sun}$/yr). This result is in line with our rather
rough estimate for the lower $\dot M$ limit ($3 \times 10^{-5}
\mathrm{M}_{\sun}$/yr).


\section{Conclusions}

We conclude that the Li abundance measured for M45 is probably to be ascribed
to the onset of rather extreme CBP phenomena on the AGB. This interpretation
would be in line with suggestions by \citet{Nollett} and is the first hint
for deep mixing coming from an M-type AGB star. As a consequence of the
efficient mixing, we expect the $^{12}$C/$^{13}$C ratio to be close to the
equilibrium value of $3.5$. A remarkable $^{14}$N enhancement at the surface
would occur only if zones of relatively high temperature ($\log T > 7.4$) were
involved in the mixing episodes. This enhancement is instead expected in any
case if HBB was at play.
We leave these predictions for a subsequent verification of the consistency
of our explanation. At this stage we can only consider CBP as the most
likely hypothesis, but we cannot derive too strong conclusions against the
other interpretations, in particular the binary wind accretion hypothesis,
on the basis of current observations.

Our findings also suggest that the conditions for CBP and those for the third
dredge up are independent. CBP is expected to be favoured at low masses, while
below a certain mass limit (depending on the metallicity) dredge-up of
s-processed material does not occur. The Li-rich stars M45 and M794 should
therefore be TP-AGB objects of really low mass ($M \le 1.3 - 1.5
\mathrm{M}_{\sun}$), where very efficient CBP can occur, while the core is not
massive enough to drive dredge-up.

\begin{figure}
  \centering
  \includegraphics[width=8.5cm]{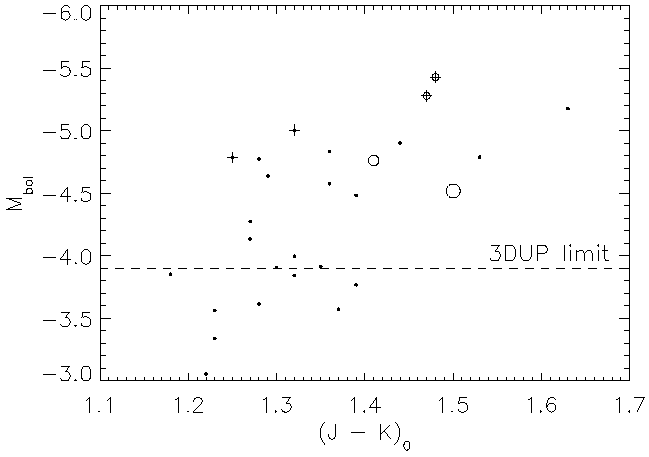} 
  \caption{
    Bolometric magnitude versus $(J - K)_{0}$ of our sample stars. Open
    circles are stars with positive Li detection, the size of the circle
    corresponds to the determined abundance of Li. Symbols with a cross denote
    stars with positive Tc detection. The dashed horizontal line indicates the
    estimated lower luminosity limit for 3DUP.
  }
  \label{MbolJK}
\end{figure}

Despite the lack of a physical explanation for the origin of CBP, a few
hints can be found from the investigation of the present homogeneous sample.
Judging from Fig.~\ref{MbolJK}, the luminosities and temperatures of the
Li-rich stars without Tc are similar to the Tc-rich stars. Also, the pulsation
periods are comparable: The Li-only stars M45 and M794 have periods of around
or just slightly below 300~d, which is about the minimum period of the Tc
stars.

M45 and M794 have Li abundances higher than the two Tc stars M1147 and M1347,
and M626 and S942, the two remaining Tc stars, do not show Li at all.
It is quite thinkable that in the mass range of our sample stars Li production
via CBP sets in before 3DUP occurs, or for masses slightly lower than
necessary to drive dredge-up. We note that the radiative layers below the
convective envelope are sub-adiabatic, chemically homogeneous zones where no
molecular weight barrier hampers the circulation of matter, so that the onset
of CBP should be much easier than formal dredge-up.

In any case, since not all of the stars with pulsation periods around or above
300~d and luminosities comparable to the Li-rich stars do show this element,
very special conditions have to be met in order for CBP to work at the rather
extreme rates necessary. Which these conditions are remains a question to be
answered.


\begin{acknowledgements}
TL and SU acknowledge funding by the Austrian Science Fund FWF under the
project P18171-N02, and BA acknowledges funding by the FWF under project
P19503-N13. MTL has been supported by the Austrian Academy of Sciences
(DOC-programme). SP and MB acknowledge funding from MURST under contract
PRIN2006-022731. We also acknowledge J. Hron (Vienna) for devising the bulge
UVES-project and for carrying out the observations.
\end{acknowledgements}


\begin{thebibliography}{}


\bibitem[Abia et al., 1991]{Abia} Abia, C., Boffin, H. M. J., Isern, J. \&
  Rebolo, R. 1991, A\&A, 245, L1

\bibitem[Bodenheimer, 1965]{Bodenheimer} Bodenheimer, P. 1965, ApJ, 142, 451

\bibitem[Boothroyd \& Sackmann, 1999]{bs99} Boothroyd, A.I., \& Sackmann,
  I.-J. 1999, ApJ, 510, 232

\bibitem[Busso et al., 1999]{Busso99} Busso, M., Gallino, R., \& Wasserburg,
  G. J. 1999, ARA\&A, 37, 239

\bibitem[Busso et al., 2003]{Busso03} Busso, M., Gallino, R., \& Wasserburg,
  G.J. 2003, PASA, 20, 356.

\bibitem[Busso et al., 2007]{Busso07} Busso, M., Wasserburg, G. J., Nollett,
  K. M., \& Calandra, A. 2007, submitted to ApJ

\bibitem[Cameron \& Fowler, 1971]{CameronFowler} Cameron, A. G. W., \& Fowler,
  W. A. 1971, ApJ, 164, 111

\bibitem[Charbonnel, 2004]{Char04} Charbonnel, C. 2004, in: McWilliam, A., \&
  Rauch, M. (Eds.), Origin and Evolution of the Elements, Cambridge
  Univ. Press, Cambridge, UK, 2004, p. 60.

\bibitem[Charbonnel \& Balachandran, 2000]{charbal} Charbonnel, C., \&
  Balachandran, S. C. 2000, A\&A, 359, 563

\bibitem[Charbonnel \& Do Nascimiento, 1998]{chana} Charbonnel, C., \& Do
  Nascimiento, J. D. 1998, A\&A, 336, 915

\bibitem[Charbonnel \& Zahn, 2007]{ChaZah} Charbonnel, C., \& Zahn,
  J.-P. 2007, A\&A, 467, 15

\bibitem[de la Reza et al., 1997]{delaReza1997} de la Reza, R., Drake, N. A.,
  \& da Silva, L. 1997, ApJ, 482, L77

\bibitem[Eggleton et al., 2006]{Eggleton} Eggleton, P. P., Dearborn, D. S. P.,
  \& Lattanzio, J. C. 2006, Science, 314, 1580
%

\bibitem[Garc\'{i}a-Hern\'{a}ndez et al., 2007]{GarciaHernandez}
  Garc\'{i}a-Hern\'{a}ndez, D. A., Garc\'{i}a-Lario, P., Plez, B., et
  al. 2007, A\&A, 462, 711

\bibitem[Gilroy \& Brown, 1991]{gil91} Gilroy, K. K., \& Brown, J. A. 1991,
  ApJ, 371, 578

%
%
%

\bibitem[Habing \& Olofsson, 2004]{HabingOlofsson} Habing, H. J., \& Olofsson,
  H. 2004, ``Asymptotic Giant Branch Stars'', A\&A Library, Springer, New York

\bibitem[Herwig, 2005]{Herwig} Herwig, F. 2005, ARAA, 43, 435

\bibitem[Iben, 1973]{Iben1973} Iben, I. 1973, ApJ, 185, 209

%
%
\bibitem[Karakas, 2003]{thesiskarakas} Karakas, A. 2003, PhD thesis, Monash
  University, Australia

\bibitem[Korn et al., 2006]{Korn} Korn, A., Grundahl, F., Richard, O., et al.
 2006, Nature, 442, 657

\bibitem[Kraft, 1994]{kra94} Kraft, R. P. 1994, PASP, 106, 553
%

\bibitem[McWilliam \& Rich, 1994]{McWilliamRich} Mcwilliam, A., \& Rich,
  R. M. 1994, ApJS, 91, 749

\bibitem[Nollett et al., 2003]{Nollett} Nollett, M. N., Busso, M., \&
  Wasserburg, G. J. 2003, ApJ, 582, 1036
%
%
%


\bibitem[Rebolo, 1991]{Rebolo} Rebolo, R. 1991, IAUS, 145, 85

\bibitem[Reiners, 2005]{Reiners} Reiners, A. 2005, AN, 326, 930
%

\bibitem[Schwenke, 1998]{Schwenke} Schwenke, D. 1998, Faraday Discuss., 109,
 321

\bibitem[Siess \& Livio, 1999]{SiessLivio} Siess, L., \& Livio, M. 1999,
  MNRAS, 304, 925

\bibitem[Smith \& Lambert, 1989]{SL1989} Smith, V. V., \& Lambert, D. L. 1989,
  ApJ, 345, L75

\bibitem[Smith \& Lambert, 1990]{SL1990} Smith, V. V., \& Lambert, D. L. 1990,
  ApJ, 361, L69

\bibitem[Smith et al., 1995]{Smith95} Smith, V. V., Plez, B., Lambert, D. L.,
  \& Lubowich, D. A. 1995, ApJ, 441, 735

\bibitem[Smith et al., 1999]{Smith1999} Smith, V.V., Shetrone, M. D., \&
  Keane, M. J. 1999, ApJ, 516, L73

\bibitem[Spite \& Spite, 1982]{Spites} Spite, M., \& Spite, F. 1982, Nature,
 297, 483

\bibitem[Straniero et al., 2003]{Stra03} Straniero, O., Dom\'{i}nguez, I.,
  Cristallo, S., \& Gallino, R. 2003, PASA, 20, 389
%
%
%

\bibitem[Uttenthaler et al., 2007]{Uttenthaler2007} Uttenthaler, S., Hron, J.,
  Lebzelter, T., et al. 2007, A\&A, 463, 251; Tc-Paper

\bibitem[Van Eck et al., 1998]{vaneck} Van Eck, S., Jorissen, A., Udry, S.,
  Mayor, M., Pernier, B. 1998, A\&A, 329, 971

\bibitem[Vanture et al., 2007]{Vanture} Vanture, A. D., Smith, V. V., Lutz,
  J., et al. 2007, PASP, 119, 147
%

\bibitem[Wasserburg et al., 1995]{Wasserburg1995} Wasserburg, G. J., Boothroyd
  A. I., \& Sackmann, I.-J. 1995, ApJ, 447, L37

\bibitem[Wesselink, 1987]{Wess87} Wesselink, Th. J. H. 1987, Ph.D. thesis,
  Catholic University of Nijmengen, the Netherlands

\end{thebibliography}
\end{document}